# Semiquantum Private Comparison of Size Relationship Based on $d$-level Single-Particle States

Mao-Jie Geng, Tian-Jie Xu, Ying Chen, Tian-Yu Ye*

College of Information & Electronic Engineering, Zhejiang Gongshang University, Hangzhou 310018, P.R.China

E-mail：happyyty@aliyun.com(T.Y.Ye)

**Abstract:** In this paper, we propose a novel semiquantum private comparison (SQPC) protocol of size relationship based on $d$-level single-particle states. The designed protocol can compare the size relationship of different privacy messages from two classical users with the help of a semi-honest third party (TP), who is permitted to misbehave on her own but cannot be in collusion with anyone else. The correctness analysis shows that this protocol can gain correct comparison results. The security analysis turns out that this protocol can resist famous outside attacks and participant attacks. Moreover, this protocol can guarantee that TP does not know the accurate comparison results. Compared with the only existing SQPC protocol of size relationship (Quantum Inf. Process. 20:124 (2021)), this protocol takes advantage over it on the aspects of initial quantum resource, TP's measurement operations and TP's knowledge about the comparison results.

## 1 Introduction

In the year of 1982, Yao [1] took the lead in proposing a classical privacy comparison protocol, which is vividly known as "the millionaire problem". Its purpose is to compare who is richer without divulging the real wealth of millionaires. Since then, classical privacy comparison has played a central role in the scenarios such as secret ballot, electronic auction, data mining and so on. The security of a classical privacy comparison protocol highly relies on the computational complexity. However, with the development of quantum parallel computing, classical privacy comparison may become more and more fragile.

In the year of 1984, Bennett and Brassard [2] put forward the first quantum cryptography protocol, i.e., the BB84 quantum key distribution (QKD) protocol. Since then, quantum cryptography has been developed rapidly. In the year of 2009, Yang and Wen [3] proposed the first quantum private comparison (QPC) protocol, which promoted classical private comparison into the quantum realm for the first time. Subsequently, scholars proposed a series of QPC protocols [4-15]. The existing QPC protocols can be classified into two kinds: the ones only comparing the equality of privacy information and the ones comparing the size relationship of privacy information. For example, the protocols in Refs. [3-9] belong to the first kind of QPC protocol, while the protocols in Refs. [10-15] belong to the second kind of QPC protocol.

In the years of 2007 and 2009, Boyer *et al.* [16, 17] proposed two representative semiquantum key distribution (SQKD) schemes, which means the birth of semiquantum cryptography. Later, in the years of 2016, Chou *et al.* [18] put forward the first semiquantum private comparison (SQPC) protocol based on Bell entangled states and quantum entanglement swapping. Subsequently, scholars proposed many SQPC protocols [19-23]. For example, Ref.[19]

put forward a measure-resend SQPC protocol by using two-particle product states to compare the equality of two classical users' secrets; Refs.[20-22] proposed different SQPC protocols by using single photons to compare the equality of two classical users' secrets; Ref.[23] designed a SQPC protocol based on $d$-dimensional Bell states, which can compare the size relationship of two classical users' secrets. It is worth noting that at present, the SQPC protocol of Ref.[23] is the only SQPC protocol which can realize the size relationship comparison of two classical users' secrets.

Based on the above analysis, in order to realize the size relationship comparison of two classical users' secret messages, we propose a new SQPC protocol of size relationship based on $d$-level single-particle states in this paper. Our protocol can compare the size relationship of secret messages from two classical users with the help of a semi-honest third party (TP). In accordance with Ref.[24], the term 'semi-honest' means that TP is allowed to misbehave on her own but cannot be in collusion with anyone else. Our protocol can guarantee that TP doesn't know the accurate comparison results.

The left parts of this paper are arranged as follows: Section 2 depicts the steps of the proposed SQPC protocol of size relationship based on $d$-level single-particle states; Section 3 conducts the correctness analysis; Section 4 validates its security in detail; and finally, Section 5 gives the discussions and conclusions.

## 2 Protocol description

In a $d$-level Hilbert space, a common orthogonal basis can be defined as $MB_Z = \{|0\rangle, |1\rangle, \ldots, |d-1\rangle\}$. After performing the $d$-level discrete quantum Fourier transform on the states within the $MB_Z$ basis, we can get the new orthogonal basis $MB_F = \{F|0\rangle, F|1\rangle, \ldots, F|d-1\rangle\}$, where

$$F|j\rangle = \frac{1}{\sqrt{d}} \sum_{k=0}^{d-1} e^{\frac{2\pi i j k}{d}} |k\rangle \tag{1}$$

for $j = 0, 1, \ldots, d-1$. Here, $MB_Z$ and $MB_F$ are two common conjugate bases.

Suppose that two classical users, Alice and Bob, have secret messages $X = (x_1, x_2, \ldots, x_n)$ and $Y = (y_1, y_2, \ldots, y_n)$, respectively, where $x_i, y_i \in \{0, 1, \ldots, h\}$, $i = 1, 2, \ldots, n$ and $d = 2h + 1$. Besides, Alice and Bob share a secret key $K_{AB} = \{K_{AB}^1, K_{AB}^2, \ldots, K_{AB}^n\}$ through a secure mediated SQKD protocol [25] in advance, where $K_{AB}^i \in \{0, 1, \cdots, d-1\}$. The specific process of our proposed SQPC protocol of size relationship based on $d$-level single-particle states is described as follows. Its flow chart is shown in Fig.1 for clarity.

Step 1: TP prepares a single-particle state sequence of length $N = 8n(1+\delta)$, $S_A = \{|S_A^1\rangle, |S_A^2\rangle, \ldots, |S_A^N\rangle\}$ ($S_B = \{|S_B^1\rangle, |S_B^2\rangle, \ldots, |S_B^N\rangle\}$), where $|S_A^i\rangle$ ($|S_B^i\rangle$) is randomly chosen from the set $MB_Z$ or $MB_F$ with equal probability. Then, TP sends the particles of $S_A$ ($S_B$) to Alice (Bob) one by one. Note that after TP transmits the first particle to Alice (Bob), she transmits a particle only after receiving the previous one.

Step 2: After receiving each particle from TP, Alice (Bob) randomly selects one of the following two operations: measuring the received particle with the $MB_Z$ basis and resending a fresh one to TP in the same state as found (this is called as the MEASURE mode); and reflecting the received particle back to TP (this is called as the REFLECT mode). Alice (Bob) writes down the corresponding measurement results when she (he) chooses to MEASURE.

Step 3: After receiving all particles sent by Alice (Bob), TP sends a confirmation signal to her (him). Then TP measures each of them in the initial prepared basis. After that, TP announces the particles which were prepared in the $MB_Z$ basis, while Alice (Bob) announces the positions where she (he) chose to REFLECT. Here, we define four kinds of particles, $B_Z$-$M$ ones, $B_Z$-$R$ ones, $B_F$-$R$ ones and $B_F$-$M$ ones, as shown in Table 1. It is worth noting that the number of each type of particles is close to $\frac{N}{4}$.

Table 1  Four kinds of particles and their applications

| Type | Initial prepared basis of TP | Alice's (Bob's) operation | Application |
| --- | --- | --- | --- |
| $B_Z$-$M$ | $MB_Z$ | MEASURE | Eavesdropping detection and private comparison |
| $B_Z$-$R$ | $MB_Z$ | REFLECT | Eavesdropping detection |
| $B_F$-$R$ | $MB_F$ | REFLECT | Eavesdropping detection |
| $B_F$-$M$ | $MB_F$ | MEASURE | Ignored |

For the $B_Z$-$R$ particles and the $B_F$-$R$ particles, TP calculates the error rates by comparing her measurement results with their initial prepared states. If both the error rate of $B_Z$-$R$ particles and the error rate of $B_F$-$R$ particles are lower than the threshold, they will continue the protocol; otherwise, they will terminate it.

Step 4: Among the $B_Z$-$M$ particles, TP randomly chooses $n$ ones as the TEST particles, and tells Alice (Bob) the chosen positions. After Alice (Bob) publishes the measurement results of these TEST particles, TP checks the error rate on them. If the error rate is lower than the threshold, they will continue the protocol; otherwise, they will terminate it.

Step 5: TP and Alice (Bob) pick out the first $n$ ones from the remaining $B_Z$-$M$ particles for private comparison. The classical values of the measurement results of these $n$ chosen particles in Alice's (Bob's) hand are represented by $S_{A1} = \{S_{A1}^1, S_{A1}^2, \ldots, S_{A1}^n\}$ ($S_{B1} = \{S_{B1}^1, S_{B1}^2, \ldots, S_{B1}^n\}$). Note that since TP prepares these $n$ chosen particles by herself, she can automatically know $S_{A1}$ ($S_{B1}$). At the same time, TP generates a classical integer sequence $S_{A2}$ ($S_{B2}$), which satisfies $S_{A1}^i \oplus S_{A2}^i = S$ ($S_{B1}^i \oplus S_{B2}^i = S$). Here, the symbol $\oplus$ represents the addition module $d$; $S$ is a constant whose value is only known by TP; and $S \in \{0,1,\ldots,d-1\}$, $S_{A1}^i, S_{A2}^i, S_{B1}^i, S_{B2}^i \in \{0,1,\ldots,d-1\}$, $i=1,2,\ldots,n$. Then, Alice (Bob) encodes her (his) secret message according to the parity of $K_{AB}$: if $\mathrm{mod}(K_{AB}^i, 2) = 0$, let $m_A^i = x_i$ ($m_B^i = y_i$); and if $\mathrm{mod}(K_{AB}^i, 2) = 1$, let $m_A^i = (d-1) - x_i$ ($m_B^i = (d-1) - y_i$). After that, Alice (Bob) calculates $R_A^i = S_{A1}^i \oplus m_A^i \oplus K_{AB}^i$ ($R_B^i = S_{B1}^i \oplus m_B^i \oplus K_{AB}^i$). Finally, Alice (Bob) sends $R_A$ ($R_B$) to TP via the classical channel, where $R_A = \{R_A^1, R_A^2, \ldots, R_A^n\}$ ($R_B = \{R_B^1, R_B^2, \ldots, R_B^n\}$).

Step 6: TP computes $R_T^i = (R_B^i \oplus S_{B2}^i) \ominus (R_A^i \oplus S_{A2}^i)$, where the symbol $\ominus$ represents the subtraction module $d$, and $i=1,2,\ldots,n$. Then, TP derives a variable $r_i$ from $R_T^i$ according to the following rule:

$$r_i = \begin{cases} 0, & if \quad R_T^i = 0; \\ -1, & if \quad 0 < R_T^i \leq h; \\ 1, & if \quad h < R_T^i \leq 2h. \end{cases} \quad (2)$$

After that, TP publishes $r_i$ to Alice and Bob.

Step 7: According to $r_i$ and $K_{AB}^i$, Alice and Bob obtain the size relationship of $x_i$ and $y_i$ by calculating the following equation, where $i = 1,2,\ldots,n$:

$$R_i = \begin{cases} 0, & if \quad r_i \times (-1)^{\mathrm{mod}(K_{AB}^i, 2)} = 0; \\ -1, & if \quad r_i \times (-1)^{\mathrm{mod}(K_{AB}^i, 2)} = -1; \\ 1, & if \quad r_i \times (-1)^{\mathrm{mod}(K_{AB}^i, 2)} = 1. \end{cases} \quad (3)$$

Here, $R_i = 0$ means $x_i = y_i$; $R_i = -1$ means $x_i < y_i$; and $R_i = 1$ means $x_i > y_i$.

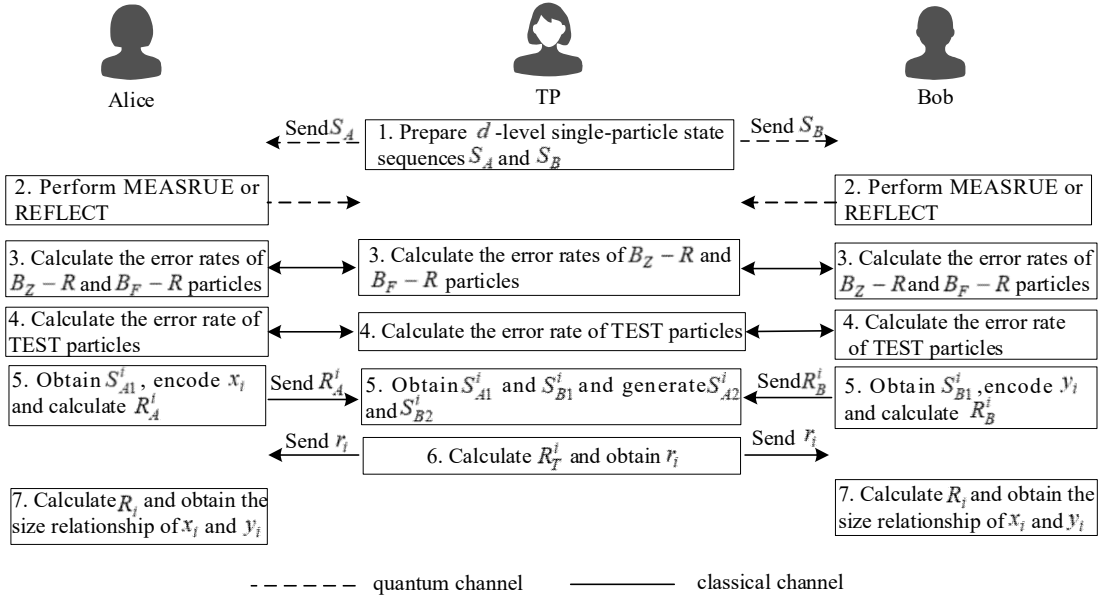

Fig. 1  The flow chart of the proposed SQPC protocol of size relationship

## 3  Correctness analysis

Here, we will analyze the correctness of the output results. According to $S_{A1}^i \oplus S_{A2}^i = S$, $S_{B1}^i \oplus S_{B2}^i = S$, $R_A^i = S_{A1}^i \oplus m_A^i \oplus K_{AB}^i$ and $R_B^i = S_{B1}^i \oplus m_B^i \oplus K_{AB}^i$, we can obtain

$$\begin{aligned} R_T^i &= \left( R_B^i \oplus S_{B2}^j \right) \ominus \left( R_A^j \oplus S_{A2}^j \right) \\ &= \left( S_{B1}^i \oplus m_B^i \oplus K_{AB}^i \oplus S_{B2}^i \right) \ominus \left( S_{A1}^i \oplus m_A^i \oplus K_{AB}^i \oplus S_{A2}^i \right) \\ &= \left( m_B^i \oplus K_{AB}^i \oplus S \right) \ominus \left( m_A^i \oplus K_{AB}^i \oplus S \right) \\ &= m_B^i \ominus m_A^i \end{aligned} \quad (4)$$

According to Eq.(4) and Alice and Bob's encoding rules, it can be obtained that if $\mathrm{mod}(K_{AB}^i, 2) = 0$, then $R_T^i = y_i \ominus x_i$; and if $\mathrm{mod}(K_{AB}^i, 2) = 1$, then $R_T^i = x_i \ominus y_i$. According to Eq.(2), we have: if $h_i^1 \ominus h_i^2 = 0$, then $r_i = 0$, which means $h_i^1 = h_i^2$; if $0 < h_i^1 \ominus h_i^2 \leq h$, then $r_i = -1$, which means $h_i^1 > h_i^2$; and if $h < h_i^1 \ominus h_i^2 \leq 2h$, then $r_i = 1$, which means $h_i^1 < h_i^2$. Here, $h_i^1, h_i^2 \in \{x_i, y_i\}$. Further, according to Eq.(3), when $\mathrm{mod}(K_{AB}^i, 2) = 0$, it has $r_i = \begin{cases} 0, & if \quad R_i = 0; \\ -1, & if \quad R_i = -1; \\ 1, & if \quad R_i = 1. \end{cases}$ and when $\text{mod}(K_{AB}^i, 2) = 1$, it has $r_i = \begin{cases} 0, & \text{if } R_i = 0; \\ -1, & \text{if } R_i = 1; \\ 1, & \text{if } R_i = -1. \end{cases}$ Consequently, it has: $R_i = 0$ means $x_i = y_i$;

$R_i = -1$ means $x_i < y_i$; and $R_i = 1$ means $x_i > y_i$. It can be concluded now that the output of our protocol is correct.

For the sake of clarity, an example is further given to verify the correctness of the comparison result. Assume that $x_i = 2$, $y_i = 4$, $h = 5$, $d = 11$ and $\text{mod}(K_{AB}^i, 2) = 1$. The classical value of the measurement result of the chosen particle for private comparison in Alice's (Bob's) hand is $S_{A1}^i$ ($S_{B1}^i$). Since TP prepares this chosen particle by herself, she can automatically know $S_{A1}^i$ ($S_{B1}^i$). TP generates a classical integer $S_{A2}^i$ ($S_{B2}^i$) to make $S_{A1}^i \oplus S_{A2}^i = S$ ($S_{B1}^i \oplus S_{B2}^i = S$). Here, $S$ is a constant and $S_{A1}^i, S_{A2}^i, S_{B1}^i, S_{B2}^i, S \in \{0,1,\ldots,d-1\}$. As $\text{mod}(K_{AB}^i, 2) = 1$, Alice (Bob) encodes $x_i$ ($y_i$) into $m_A^i = (d-1) - x_i = 8$ ($m_B^i = (d-1) - y_i = 6$). After that, Alice (Bob) calculates $R_A^i = S_{A1}^i \oplus m_A^i \oplus K_{AB}^i$ ($R_B^i = S_{B1}^i \oplus m_B^i \oplus K_{AB}^i$). Finally, Alice (Bob) sends $R_A^i$ ($R_B^i$) to TP. TP computes $R_T^i = (R_B^i \oplus S_{B2}^i) \ominus (R_A^i \oplus S_{A2}^i) = m_B^i \ominus m_A^i = 9$. According to Eq.(2), TP gets $r_i = 1$. TP publishes $r_i$ to Alice and Bob. As $r_i \times (-1)^{\text{mod}(K_{AB}^i, 2)} = -1$, according to Eq.(3), Alice and Bob obtain $R_i = -1$. As a result, Alice and Bob conclude that $x_i < y_i$.

## 4 Security analysis

In this section, we validate that the proposed protocol is secure against both the outside attacks and the participant attacks.

### 4.1 Outside attacks

In the proposed protocol, Alice plays the similar role to Bob. Without loss of generality, we only validate the transmission security of single particles from TP to Alice and back to TP. During these transmissions, Eve may launch some famous attacks, such as the intercept-resend attack, the measure-resend attack and the entangle-measure attack, to get something useful about Alice's secret message.

(1) The intercept-resend attack

For trying to know Alice's measurement result on the particle for private comparison, Eve intercepts the particle of $S_A$ from TP to Alice and sends the fake one she generated beforehand in the $MB_Z$ basis to Alice; after Alice executes her operation on the fake particle, Eve intercepts the particle returned by Alice and sends the genuine one in her hand to TP. Apparently, if Alice chooses to REFLECT, Eve's attack will not be detected in Step 3, no matter what the genuine particle is. Then, consider the case that Alice chooses to MEASURE. If the genuine particle is in the $MB_F$ basis, according to Table 1, this particle is ignored so that Eve's attack will not be detected; and if it is in the $MB_Z$ basis and chosen as a TEST particle, Eve's attack will be detected with the probability of $\frac{d-1}{d}$ in Step 4. To sum up, the reason why Eve's this kind of attack can be detected inevitably lies in two aspects: on the one hand, Alice's operation is random to Eve; and on the other hand, Eve's fake particle is likely to be different from the genuine one prepared by TP.

(2) The measure-resend attack

In order to try to know Alice's measurement result on the particle for private comparison, Eve intercepts the particle of $S_A$ from TP to Alice, measures it with the $MB_Z$ basis and sends the resulted state to Alice. If this original particle is in the $MB_Z$ basis, Eve's attack will not be detected in Steps 3 and 4, no matter what operation Alice chooses. Then, consider the case that this original particle is in the $MB_F$ basis. If Alice chooses to MEASURE, according to Table 1, this particle is ignored so that Eve's attack behavior will not be detected; and if Alice chooses to REFLECT, Eve's attack behavior will be detected in Step 3, as Eve's measurement operation destroys the original state of this particle. To sum up, Eve's this kind of attack can be detected inevitably due to the following two reasons: on the one hand, Eve has no idea about the prepared basis for this particle; and on the other hand, Alice's operation is random to Eve.

In addition, for trying to know Alice's measurement result on the particle for private comparison, Eve may launch another measure-resend attack: she intercepts the particle from Alice to TP, measures it with the $MB_Z$ basis and sends the resulted state to TP. Same to the first kind of measure-resend attack, Eve is inevitably detected in this case.

(3) The entangle-measure attack

Eve's this kind of attack can be simulated by two unitaries $\hat{E}$ and $\hat{F}$. Here, $\hat{E}$ attacks the qudit from TP to Alice, while $\hat{F}$ attacks the qudit back from Alice to TP. Moreover, $\hat{E}$ and $\hat{F}$ share a common probe space with the initial state $|\varepsilon\rangle$. Just as illustrated in Refs.[16,17], the shared probe allows Eve to attack the returned particle by taking advantage of the knowledge acquired from $\hat{E}$ (if Eve does not make use of this, the 'shared probe' can simply be considered as the composite system formed by two independent probes). Any attack where Eve would let $\hat{F}$ rely on a measurement after implementing $\hat{E}$ can be accomplished by $\hat{E}$ and $\hat{F}$ with controlled gates. Eve's entangle-measure attack on the particle of $S_A$ can be described as Fig.2.

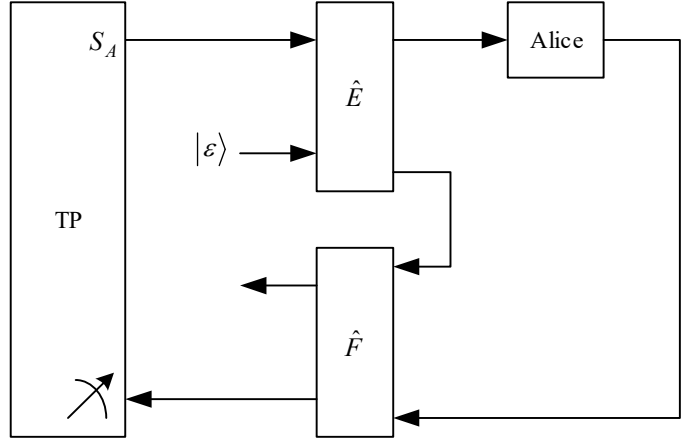

Fig.2　Eve's entangle-measure attack on the particle of $S_A$ with two unitaries $\hat{E}$ and $\hat{F}$

**Theorem 1.** *Suppose that Eve performs attack $(\hat{E}, \hat{F})$ on the particle from TP to Alice and back to TP. For this attack inducing no error in Steps 3 and 4, the final state of Eve's probe should be independent of not only Alice's operation but also TP and Alice's measurement results. As a result, Eve gets no information on the values of $S_{A1}$.*

**Proof.** The global state of the composite system consisted of the qudit prepared by TP and Eve's probe before Eve's attack can be expressed as $|T\rangle|\varepsilon\rangle$, where $|T\rangle$ is randomly chosen from the two sets $MB_Z$ and $MB_F$. For convenience, we use $|g\rangle$ and $|G_g\rangle$ to represent the particle in the set $MB_Z$ and the particle in the set $MB_F$, respectively, where $|G_g\rangle = F|g\rangle = \frac{1}{\sqrt{d}}\sum_{k=0}^{d-1}e^{\frac{2\pi igk}{d}}|k\rangle$ and $g = 0,1,\ldots,d-1$.

①Firstly, consider the case that the particle sent from TP to Alice is in the set $MB_Z$. The effect of $\hat{E}$ on this particle and Eve's probe can be described as [26]

$$\hat{E}(|0\rangle|\varepsilon\rangle) = \alpha_{00}|0\rangle|\varepsilon_{00}\rangle + \alpha_{01}|1\rangle|\varepsilon_{01}\rangle + \cdots + \alpha_{0(d-1)}|d-1\rangle|\varepsilon_{0(d-1)}\rangle, \tag{5}$$

$$\hat{E}(|1\rangle|\varepsilon\rangle) = \alpha_{10}|0\rangle|\varepsilon_{10}\rangle + \alpha_{11}|1\rangle|\varepsilon_{11}\rangle + \cdots + \alpha_{1(d-1)}|d-1\rangle|\varepsilon_{1(d-1)}\rangle, \tag{6}$$

$$\vdots$$

$$\hat{E}(|d-1\rangle|\varepsilon\rangle) = \alpha_{(d-1)0}|0\rangle|\varepsilon_{(d-1)0}\rangle + \alpha_{(d-1)1}|1\rangle|\varepsilon_{(d-1)1}\rangle + \cdots + \alpha_{(d-1)(d-1)}|d-1\rangle|\varepsilon_{(d-1)(d-1)}\rangle, \tag{7}$$

where $|\varepsilon_{gt}\rangle$ is Eve's probe states determined by $\hat{E}$, and $g,t = 0,1,\ldots,d-1$; and for $g = 0,1,\ldots,d-1$, there is $\sum_{t=0}^{d-1}|\alpha_{gt}|^2 = 1$.

After Alice's operation, Eve imposes $\hat{F}$ on the particle sent back to TP.

When Alice chooses to MEASURE, according to the equations from (5) to (7), the global system of the composite system is collapsed into $\alpha_{gt}|t\rangle|\varepsilon_{gt}\rangle$, where $g,t = 0,1,\ldots,d-1$. In order that Eve's attacks on the particle will not be discovered by TP and Alice in Step 4, the global state of the composite system after Eve imposes $\hat{F}$ should be involved into

$$\hat{F}(\alpha_{gt}|t\rangle|\varepsilon_{gt}\rangle) = \begin{cases} \alpha_{gg}|g\rangle|F_{gg}\rangle, & \text{if } g = t; \\ 0, & \text{if } g \neq t. \end{cases} \tag{8}$$

which includes

$$\alpha_{gt} = \begin{cases} \alpha_{gg}, & \text{if } g = t; \\ 0, & \text{if } g \neq t. \end{cases} \tag{9}$$

When Alice chooses to REFLECT, according to Eq.(8) and Eq.(9), the global state of the composite system after Eve imposes $\hat{F}$ is involved into

$$\hat{F}[\hat{E}(|0\rangle|\varepsilon\rangle)] = \hat{F}(\alpha_{00}|0\rangle|\varepsilon_{00}\rangle + \alpha_{01}|1\rangle|\varepsilon_{01}\rangle + \cdots + \alpha_{0(d-1)}|d-1\rangle|\varepsilon_{0(d-1)}\rangle) = \hat{F}(\alpha_{00}|0\rangle|\varepsilon_{00}\rangle) = \alpha_{00}|0\rangle|F_{00}\rangle, \tag{10}$$

$$\hat{F}[\hat{E}(|1\rangle|\varepsilon\rangle)] = \hat{F}(\alpha_{10}|0\rangle|\varepsilon_{10}\rangle + \alpha_{11}|1\rangle|\varepsilon_{11}\rangle + \cdots + \alpha_{1(d-1)}|d-1\rangle|\varepsilon_{1(d-1)}\rangle) = \hat{F}(\alpha_{11}|1\rangle|\varepsilon_{11}\rangle) = \alpha_{11}|1\rangle|F_{11}\rangle, \tag{11}$$

$$\vdots$$

$$\hat{F}[\hat{E}(|d-1\rangle|\varepsilon\rangle)] = \hat{F}(\alpha_{(d-1)0}|0\rangle|\varepsilon_{(d-1)0}\rangle + \alpha_{(d-1)1}|1\rangle|\varepsilon_{(d-1)1}\rangle + \cdots + \alpha_{(d-1)(d-1)}|d-1\rangle|\varepsilon_{(d-1)(d-1)}\rangle) = \hat{F}(\alpha_{(d-1)(d-1)}|d-1\rangle|\varepsilon_{(d-1)(d-1)}\rangle) = \alpha_{(d-1)(d-1)}|d-1\rangle|F_{(d-1)(d-1)}\rangle. \tag{12}$$

In order that Eve's attacks on the particle will not be discovered by TP and Alice in Step 3, Eve cannot change its state when Alice chooses to REFLECT. It is automatically satisfied according to the equations from (10) to (12).

②Secondly, consider the case that the particle from TP to Alice is in the set $MB_F$. The effect of $\hat{E}$ on this particle and Eve's probe can be described as

$$\hat{E}(|G_g\rangle|\varepsilon\rangle) = \hat{E}\left[\left(\frac{1}{\sqrt{d}}\sum_{k=0}^{d-1}e^{\frac{2\pi igk}{d}}|k\rangle\right)|\varepsilon\rangle\right] = \frac{1}{\sqrt{d}}\sum_{k=0}^{d-1}e^{\frac{2\pi igk}{d}}\hat{E}(|k\rangle|\varepsilon\rangle). \tag{13}$$

After Alice's operation, Eve imposes $\hat{F}$ on the particle sent back to TP.

When Alice chooses to REFLECT, the global state of the composite system after Eve executes $\hat{F}$ is changed into

$$\hat{F}\left[\hat{E}\left(|G_g\rangle|\varepsilon\rangle\right)\right] = \frac{1}{\sqrt{d}}\sum_{k=0}^{d-1} e^{\frac{2\pi i g k}{d}} \hat{F}\left[\hat{E}\left(|k\rangle|\varepsilon\rangle\right)\right]. \tag{14}$$

On the basis of the equations from (10) to (12), we have $\hat{F}\left[\hat{E}\left(|k\rangle|\varepsilon\rangle\right)\right] = \alpha_{kk}|k\rangle|F_{kk}\rangle$. Inserting it into Eq.(14) produces

$$\hat{F}\left[\hat{E}\left(|G_g\rangle|\varepsilon\rangle\right)\right] = \frac{1}{\sqrt{d}}\sum_{k=0}^{d-1} e^{\frac{2\pi i g k}{d}} \alpha_{kk}|k\rangle|F_{kk}\rangle. \tag{15}$$

According to the inverse quantum Fourier transform, it has

$$|k\rangle = \frac{1}{\sqrt{d}}\sum_{j=0}^{d-1} e^{-\frac{2\pi i j k}{d}} |G_j\rangle, \quad k = 0,1,\ldots,d-1. \tag{16}$$

Inserting Eq.(16) into Eq.(15) produces

$$\hat{F}\left[\hat{E}\left(|G_g\rangle|\varepsilon\rangle\right)\right] = \frac{1}{\sqrt{d}}\sum_{k=0}^{d-1} e^{\frac{2\pi i g k}{d}} \alpha_{kk}\left(\frac{1}{\sqrt{d}}\sum_{j=0}^{d-1} e^{-\frac{2\pi i j k}{d}}|G_j\rangle\right)|F_{kk}\rangle$$

$$= \frac{1}{d}\left(|G_0\rangle\sum_{k=0}^{d-1} e^{\frac{2\pi i (g-0)k}{d}} \alpha_{kk}|F_{kk}\rangle + |G_1\rangle\sum_{k=0}^{d-1} e^{\frac{2\pi i (g-1)k}{d}} \alpha_{kk}|F_{kk}\rangle\right.$$

$$\left. + \cdots + |G_{(d-1)}\rangle\sum_{k=0}^{d-1} e^{\frac{2\pi i [g-(d-1)]k}{d}} \alpha_{kk}|F_{kk}\rangle\right). \tag{17}$$

In order that Eve will not be discovered by TP and Alice in Step 3, it should meet

$$\sum_{k=0}^{d-1} e^{\frac{2\pi i (g-j)k}{d}} \alpha_{kk}|F_{kk}\rangle = 0 \tag{18}$$

for $g \neq j$ and $g, j = 0,1,\ldots,d-1$. Apparently, for any $g \neq j$, it has

$$\sum_{k=0}^{d-1} e^{\frac{2\pi i (g-j)k}{d}} = 0. \tag{19}$$

Therefore, according to Eq.(18) and Eq.(19), we have

$$\alpha_{00}|F_{00}\rangle = \alpha_{11}|F_{11}\rangle = \cdots = \alpha_{(d-1)(d-1)}|F_{(d-1)(d-1)}\rangle = \alpha|F\rangle. \tag{20}$$

③Inserting Eq.(20) into Eq.(8) produces

$$\hat{F}\left(\alpha_{gt}|t\rangle|\varepsilon_{gt}\rangle\right) = \begin{cases} |g\rangle(\alpha|F\rangle), & \text{if } g = t; \\ 0, & \text{if } g \neq t. \end{cases} \tag{21}$$

Inserting Eq.(20) into the equations from (10) to (12) produces

$$\hat{F}\left[\hat{E}\left(|0\rangle|\varepsilon\rangle\right)\right] = |0\rangle(\alpha|F\rangle), \tag{22}$$

$$\hat{F}\left[\hat{E}\left(|1\rangle|\varepsilon\rangle\right)\right] = |1\rangle(\alpha|F\rangle), \tag{23}$$

$$\vdots$$

$$\hat{F}\left[\hat{E}\left(|d-1\rangle|\varepsilon\rangle\right)\right] = |d-1\rangle(\alpha|F\rangle), \tag{24}$$

respectively. Applying Eq.(20) into Eq.(17) generates

$$\hat{F}\left[\hat{E}\left(|G_g\rangle|\varepsilon\rangle\right)\right] = |G_g\rangle(\alpha|F\rangle). \tag{25}$$

Based on the equations from (21) to (25), it can be concluded that for Eve not inducing an error in Steps 3 and 4, the final state of Eve's probe should be independent of not only Alice's operations but also TP and Alice's measurement results. So, if Eve launches an entangle-measure attack, she will not get the information on the values of $S_{A1}$.

(4) The Trojan horse attack

The particles of $S_A$ go a round trip, as they are sent from TP to Alice and back to TP, so we need to consider the Trojan horse attack launched by Eve, such as the invisible photon eavesdropping attack [27] and the delay-photon Trojan horse attack [28,29]. The approach of resisting the invisible photon eavesdropping attack is that Alice puts a wavelength filter in front of her device to erase the illegitimate photon signal [29,30]. The method of preventing the delay-photon Trojan horse attack is that Alice uses a photon number splitter (PNS: 50/50) to split each sample signal into two parts and utilizes the correct measuring bases to measure the resulted signals [29,30]. This attack is discovered as long as the multiphoton rate is unreasonably high.

## 4.2 Participant attacks

In 2007, Gao *et al.* [31] first proposed that the attack launched by a dishonest participant is always more serious than that from an outside eavesdropper. Hence, we need to pay special attention to participant attacks. With respect to the proposed protocol, we need to consider two cases of participant attacks, i.e., one from Alice or Bob and the other from the semi-honest TP.

(1) The participant attack from Alice or Bob

In the proposed protocol, Alice's role is same to Bob's. Without loss of generality, we only consider that Bob, who is supposed to have complete quantum abilities, is dishonest. As a result, dishonest Bob tries his best to obtain Alice's secret message.

When Alice sends $R_A^i$ to TP via the classical channel, Bob may hear of it. Although Bob automatically knows $K_{AB}^i$, in order to deduce $m_A^i$ from $R_A^i$, he still needs to obtain $S_{A1}^i$. However, Bob cannot collude with TP to get $S_{A1}^i$. Therefore, in order to get it, Bob has to launch the active attacks on the particles of $S_A$. However, just as analyzed above, he is inevitably detected as an outside attacker by Alice and TP. In conclusion, Bob has no chance to obtain $m_A^i$. It is equivalent to say that Bob has no access to $x_i$.

(2) The participant attack from the semi-honest TP

In this protocol, TP is assumed to be semi-honest, which means that she may misbehave but cannot collude with Alice or Bob. Although TP knows $S_{A1}^i$ ($S_{B1}^i$), she still cannot derive $m_A^i$ ($m_B^i$) from $R_A^i$ ($R_B^i$), due to lack of $K_{AB}^i$. As a result, TP has no chance to get $x_i$ ($y_i$).

## 5 Discussions and conclusions

As the SQPC protocol of Ref.[23] is the only existing SQPC protocol which can realize the comparison of size relationship of private messages from two users, we compare the proposed SQPC protocol with it in detail. The comparison results are listed in Table 2. Ref. [32] defines the qubit efficiency to evaluate the efficiency of a quantum communication protocol in the 2-level system. Here, in order to calculate the efficiency of a quantum communication protocol in the $d$-level system, we modify it into the qudit efficiency, which is defined as follows: $\eta = \frac{b}{q+c}$, where $b$, $q$ and $c$ are the length of compared private messages, the number of consumed qudits and the length of classical information involved in the classical communication, respectively. Note that the classical resources used in security check processes are ignored here. Moreover, the SQPC protocol of Ref.[23] doesn't illustrate which semiquantum key agreement (SQKA) protocol is used for establishing the pre-shared key between two classical users, so the quantum resources and the classical resources consumed for generating the pre-shared key are also ignored.

In the proposed protocol, the length of Alice's (Bob's) private message is $n$, so it has $b=n$. TP needs to prepare two groups of $8n(1+\delta)$ initial $d$-level single particles and send them to Alice and Bob, respectively. Then, when Alice (Bob) chooses to MEASURE, she (he) generates $4n(1+\delta)$ fresh qudits in the $MB_Z$ basis and sends them to TP. As a result, it has $q=8n(1+\delta)\times 2+4n(1+\delta)\times 2=24n(1+\delta)$. In addition, Alice (Bob) needs to send $R_A^i$ ($R_B^i$) to TP, while TP needs to publish $r_i$ to Alice and Bob, where $i=1,2,\ldots,n$. Hence, it has $c=n+n+n=3n$. Therefore, the qudit efficiency of the proposed protocol is $\eta=\dfrac{n}{24n(1+\delta)+3n}=\dfrac{1}{24(1+\delta)+3}$.

In the SQPC protocol of Ref.[23], the length of Alice's (Bob's) private message is $n$, so we have $b=n$. TP needs to prepare $8n(1+\delta)$ initial $d$-level Bell entangled states and send them to Alice and Bob, respectively. Then, when Alice (Bob) chooses to MEASURE, she (he) prepares $4n(1+\delta)$ fresh qudits in the $\bar{Z}$ basis and transmits them to TP. Hence, we have $q=8n(1+\delta)\times 2+4n(1+\delta)\times 2=24n(1+\delta)$. In addition, Alice (Bob) needs to send $r_A^i$ ($r_B^i$) to TP, while TP needs to publish $c_r^i$ to Alice and Bob, where $i=1,2,\ldots,n$. Hence, we have $c=n+n+n=3n$. Consequently, the qudit efficiency of the SQPC protocol of Ref.[23] is $\eta=\dfrac{n}{24n(1+\delta)+3n}=\dfrac{1}{24(1+\delta)+3}$.

According to Table 2, it can be concluded that: (1) our protocol exceeds the SQPC protocol of Ref.[23] in initial quantum resource, as the preparation of $d$-level single-particle state is much easier to realize than that of $d$-level Bell entangled states; (2) our protocol takes advantage over the SQPC protocol of Ref.[23] in TP's measurement operation, as the measurement on $d$-level single-particle state is much easier to realize than that on $d$-level Bell entangled states; and (3) our protocol has better privacy for the comparison results than the SQPC protocol of Ref.[23], as TP cannot know the comparison results in our protocol.

Table 2  Comparison results of our SQPC protocol and the SQPC protocol of Ref.[23]

| | Initial quantum resource | Usage of quantum entanglement or unitary operation | Type of TP | TP's measurement operation | Usage of SQKD or SQKA | Comparison of size relationship | TP's knowledge about the comparison result | Qudit efficiency |
|---|---|---|---|---|---|---|---|---|
| Ref.[23] | $d$-level Bell entangled states | No | semi-honest | $d$-level Bell entangled state measurements and $d$-level single-particle measurements | Yes | Yes | Yes | $\dfrac{1}{24(1+\delta)+3}$ |
| Our protocol | $d$-level single-particle states | No | semi-honest | $d$-level single-particle measurements | Yes | Yes | No | $\dfrac{1}{24(1+\delta)+3}$ |

In conclusion, this paper proposes a novel SQPC protocol of size relationship based on $d$-level single-particle states, which can safely compare the size relationship of private inputs from two classical users. Our protocol can overcome a variety of outside and participant attacks and ensure that the semi-honest TP has no knowledge about the specific comparison results. Our protocol only employs $d$-level single-particle states as initial quantum resource and requires TP to perform $d$-level single- particle measurements. As a result, our protocol exceeds the only existing SQPC protocol of size relationship in Ref.[23] with respect to initial quantum resource, TP's measurement operations and TP's knowledge about the comparison results.

**Acknowledgments**

Funding by the National Natural Science Foundation of China (Grant No.62071430 and No.61871347), the Fundamental Research Funds for the Provincial Universities of Zhejiang (Grant No.JRK21002) and Zhejiang Gongshang University, Zhejiang Provincial Key Laboratory of New Network Standards and Technologies (No. 2013E10012) is gratefully acknowledged.